\documentclass[aps,
               prl,
               preprint,
               amsmath,
               amssymb,
               showpacs,
               superscriptaddress]{revtex4-1}
\usepackage{graphicx} 
\usepackage{dcolumn}  
\usepackage{bm}       
\usepackage{amsfonts}
\usepackage{dsfont}

\begin{document}

\title{Proposed imaging of the ultrafast electronic motion in samples using x-ray phase-contrast}

\author{Gopal Dixit}
\email[]{gopal.dixit@cfel.de}
\affiliation{%
Center for Free-Electron Laser Science, DESY,
            Notkestrasse 85, D-22607 Hamburg, Germany }

\author{Jan Malte Slowik}
\email[]{jan.malte.slowik@cfel.de}
\affiliation{%
Center for Free-Electron Laser Science, DESY,
            Notkestrasse 85, D-22607 Hamburg, Germany }
\affiliation{%
Department of Physics, University of Hamburg, D-20355 Hamburg,
Germany}

\author{Robin Santra}
\email[]{robin.santra@cfel.de}
\affiliation{%
Center for Free-Electron Laser Science, DESY,
            Notkestrasse 85, D-22607 Hamburg, Germany }
\affiliation{%
Department of Physics, University of Hamburg, D-20355 Hamburg, Germany}

\date{\today}



\begin{abstract}
Tracing the motion of electrons has enormous relevance 
to understanding ubiquitous phenomena in ultrafast science, such as the dynamical evolution of the electron density during 
complex chemical and biological processes. 
Scattering of ultrashort \mbox{x-ray} pulses from an electronic wavepacket
would appear to be the most obvious approach to image the electronic motion in real-time and real-space
with the notion that such scattering patterns, in the far-field regime, encode
the instantaneous electron density of the wavepacket.
However, recent results by Dixit {\em et al.} [Proc. Natl. Acad. Sci. U.S.A., {\bf 109},
11636 (2012)] have put this notion into question and shown that the 
scattering in the far-field regime probes spatio-temporal density-density correlations. 
Here, we propose a possible way to image the instantaneous electron
density of the wavepacket via ultrafast \mbox{x-ray} {\em phase contrast imaging}. 
Moreover, we show that inelastic scattering processes, which plague ultrafast scattering in the far-field regime, 
do not contribute in ultrafast \mbox{x-ray} phase contrast imaging as
a consequence of an interference effect. 
We illustrate our general findings by means of a wavepacket that lies in the time and energy range of the dynamics of valence electrons in complex molecular
and biological systems. 
This present work offers a potential to image not only instantaneous snapshots of non-stationary electron dynamics, 
but also the Laplacian of these snapshots which provide information about the complex bonding and topology of the charge distributions in the systems. 
\end{abstract}

\maketitle 
In the past, momentum spectroscopy was used to study the internal dynamics of electrons in stationary states 
~\cite{cooper1999, mccarthy1999, brion2001}.  According to quantum mechanics, the outcome of measurements 
on stationary states is time-independent. In contrast,  electronic wavepackets are non-stationary states 
which undergo time-evolution and thus give rise to time-dependent measurements. Therefore, access to the dynamics of 
non-stationary electrons provides new information about phenomena at the microscopic level.
The distinctive timescale of electronic motion,
responsible for chemical bonding and electron transfer processes
in molecules and complex biological systems, ranges from a few
femtoseconds (1 fs = 10$^{-15}$ s) to several attoseconds (1 as =
10$^{-18}$ s).
In order to gain insight into ultrafast 
chemical or physical processes, one has to unravel the
motion of electrons with spatial and temporal
resolutions of order 1~{\AA} and 1~fs, respectively~\cite{Krausz, bucksbaum2007, corkum2007}.
Pump-probe experiments are the most direct approach to investigate fast-evolving
microscopic processes, where first
a pump pulse triggers the dynamics and then subsequently a probe
pulse interrogates such triggered dynamics as a function of
pump-probe delay time. In the last few years, 
with the availability of laser pulses on the sub-fs timescale ~\cite{Goulielmakis1,
Hentschel}, remarkable progress has been made towards the understanding of electronic motion 
in real time~\cite{Haessler, Tzallas, Hockett, Niikura, Goulielmakis, Sansone, smirnova2009}.
In recent years, due to advancement in technology, it has become possible to generate 
ultraintense, ultrashort and tunable \mbox{x-ray} pulses from novel light sources such as 
free-electron lasers (FEL)~\cite{emma2, ishikawa2012}, laser plasmas ~\cite{rousse2001}
and high-harmonic generation~\cite{mckinnie2010, popmintchev2012}, which may provide a unique opportunity to
investigate these ultrafast processes with atomic-scale spatial
and temporal resolution.
Since the beginning of the operation of the first FEL in the hard \mbox{x-ray}
regime, the Linac Coherent Light Source, 
high-intensity x-ray experiments have been carried out for systems ranging from
atoms~\cite{young2010}, small molecules~\cite{berrah2011}, complex biomolecules~\cite{chapman},
to matter in extreme conditions~\cite{vinko2012}.

To image electronic motion in real-time and real-space, which is important to understand several 
ultrafast complex processes, one can perform far-field scattering 
of ultrashort \mbox{x-ray} pulses from the dynamically evolving electronic system. 
By varying the pump-probe time delay one obtaines a series of scattering patterns 
that serve to image the electronic motion with atomic-scale 
spatio-temporal resolution. Based on experience with elastic scattering  
from electronically stationary targets, one might expect to be able to retrieve 
the instantaneous electron density (IED) of the electronic wavepacket via 
ultrafast scattering in far-field regime.  However, for probing the motion of the wavepacket on
an ultrafast timescale, one needs an ultrashort pulse with
unavoidable bandwidth. Due to the inherent bandwidth, there is no way to know whether 
the probe pulse induces transitions among the eigenstates spanning
the wavepacket, or to other states closer in energy than the
bandwidth. As a result, the probe pulse
inevitably changes the wavepacket and the pattern contains
contributions from all the states within the bandwidth, that were
accessed during the scattering process. 
Therefore, it is impossible to transfer the concepts underlying stationary coherent 
scattering to the ultrafast regime. Recently, Dixit et
al.~\cite{dixit2012} have investigated ultrafast scattering
from an electronic wavepacket in the far-field regime and shown that the scattering patterns
encode spatial and temporal correlations that deviate completely from the
notion of the IED as the key quantity being probed~\cite{Krausz,
Starace}. 
Moreover, they have shown that, ultimately this is a consequence of quantum electrodynamics and 
can not be captured semiclassically~\cite{dixit2012}. 
Therefore, it is not possible to retrieve direct information about the dynamical structural changes 
during the complex processes from these correlations. On the other hand, if one can retrieve the IED at 
different instants of time during the processes, it will provide direct insight about these processes.
In light of this, one may wonder if there is any way, using light, 
to image the IED of the wavepacket in real-space at different instants of time during
the motion of the wavepacket.

\begin{figure}[ht]
\includegraphics[width=8cm]{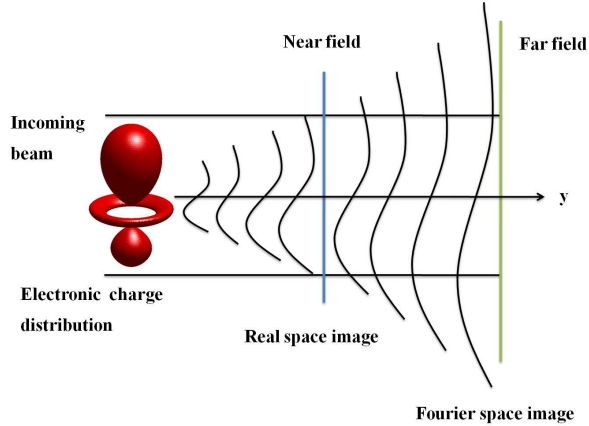}
\caption{(color online). Concept of radiation-matter interaction,
where matter is represented by an isosurface of an electronic
charge distribution of an electronic wavepacket. The
real-space phase-contrast image is observed in the near-field region, 
whereas the Fourier-space scattering pattern is observed in the far-field region.} 
\label{fig11}
\end{figure}

In this Letter, we demonstrate rigorously how the IED
of an electronic wavepacket can be imaged via ultrafast phase contrast imaging (PCI).
The two important classes of imaging are near-field imaging and
far-field imaging (see Fig.~\ref{fig11})~\cite{Nielsen, nugent2010}. 
In the following, we discuss the pros and cons of imaging
the motion of the wavepacket in both regimes.
In far-field imaging, one only detects the scattered
radiation and obtains the Fourier space image of the object from
which the scattering has taken place. 
This is what is normally done in x-ray crystallography~\cite{Nielsen}. 
On the other hand, in near-field imaging, one detects the
interference between incident and scattered radiation~\cite{Nielsen}.
\mbox{X-ray} PCI falls into the near-field imaging regime, with
many applications in various fields of science including
biology~\cite{davis1995nature, wilkins1996} and condensed matter
physics~\cite{gureyev2001, de2008}.
Photoelectron holography~\cite{chelkowski2006, huismans2011, bian2012} is an alternative approach to imaging electronic motion. 
We expect that the technique of ultrafast x-ray PCI proposed here will be particularly useful for studying spatially extended systems, 
which are often inaccessible to methods based on photoelectrons.

We apply a consistent quantum theory for the radiation field 
and the electrons, by treating both using the language of quantum field theory.
Following from the principle of minimal coupling in Coulomb gauge, the Hamiltonian 
describing the interaction between radiation field and electrons is~\cite{craig1984}
\begin{eqnarray}\label{eq1}
\hat{H}_{\textrm{int}} & = & -\frac{e}{m} \int d^{3}x
~\hat{\psi}^{\dagger}(\mathbf{x})~
\left[\hat{\mathbf{A}}(\mathbf{x}) \cdot
\frac{\hbar}{i} \mathbf{\nabla}\right] ~\hat{\psi}(\mathbf{x})
\nonumber \\
& & + \frac{e^{2}}{2~m}\int d^{3}x
~\hat{\psi}^{\dagger}(\mathbf{x})~ \hat{\mathbf{A}}^{2}
(\mathbf{x})~\hat{\psi}(\mathbf{x}).
\end{eqnarray}
Here, the field operator $\hat{\psi}^{\dagger}(\mathbf{x}) ~ [\hat{\psi}(\mathbf{x})]$
creates (annihilates) an electron at position $\mathbf{x}$; $e$ and $m$ are charge and 
mass of the electron, respectively; $(\hbar/i) \mathbf{\nabla}$ represents the canonical 
momentum of an electron, $\hbar$ being the reduced Planck constant;  
and $\hat{\mathbf{A}}$ is the vector potential operator of the
radiation field. 
Here, we only focus on scattering events induced
by the $\hat{\mathbf{A}}^{2} $ operator and will not consider the
contribution from the $\hat{\mathbf{A}}(\mathbf{x}) \cdot \mathbf{\nabla}$ term in Eq.~(\ref{eq1}) in the scattering
process as it is known that its contribution
is negligibly small at photon energies much higher than
all inner-shell thresholds in the system of
interest~\cite{Nielsen}.
However, it is useful to mention that the scattering induced by the $\hat{\mathbf{A}}(\mathbf{x}) \cdot \mathbf{\nabla}$ term 
in second order gives rise to the Kramers-Heisenberg cross section, which describes the powerful photon-in/photon-out 
technique of resonant inelastic x-ray scattering~\cite{ma1994RIXS, mills1997RIXS, schulke2007electron}.
Note that inelastic and elastic scattering at high photon energy, i.e., Compton and Thomson scattering, is mediated by the 
$\hat{\mathbf{A}}^{2} $ operator. 
Since we are working in a particular gauge (the Coulomb gauge), our approach is not gauge invariant.

Let us assume that with the help of a pump pulse, one prepares a
coherent superposition of eigenstates (electronic wavepacket) with
$\hat{\rho}^{\mathrm{el}}_{\textrm{in}} = |\Psi_{\textrm{in}}
\rangle \langle \Psi_{\textrm{in}} |$ as the initial density
operator for the wavepacket. The multimode radiation field is
treated as a collection of photons occupying different electromagnetic 
modes with $\hat{\rho}^{X}_{\textrm{in}}$ as the
initial density operator of the field~\cite{mandel1995}. Before
the interaction takes place, the entire system is therefore
prepared in the initial state $\hat{\rho}_{\textrm{in}} =
\hat{\rho}_{\textrm{in}}^{X} \otimes
\hat{\rho}_{\textrm{in}}^{\mathrm{el}}$. 
The density
operator evolves according to $\hat{\rho}(t) = \lim_{t_{0} \to -\infty}\hat{U}_{I}(t,
t_{0})~\hat{\rho}_{\textrm{in}}~\hat{U}_{I}^{\dagger}(t,t_{0})$,
with the time-evolution operator $\hat{U}_{I}$ satisfying the
equation of motion with respect to $\hat{H}_{\mathrm{int}}$ in the
interaction picture.
Making a perturbation
expansion of the time evolution operator 
to first order with respect to $\hat{H}_{\mathrm{int}}$,
the expectation value of an observable $\mathcal{\hat{O}}$ at time
$t$ is expressed as 
\begin{eqnarray}\label{eq3}
\langle \mathcal{\hat{O}} \rangle_t & = &
\textrm{Tr} \left(\hat{\rho}(t) \mathcal{\hat{O}}(t)\right) \nonumber \\
& = & \textrm{Tr} \left( \hat{\rho}_{\textrm{in}}
\mathcal{\hat{O}}(t) \right) + 2 \textrm{Re} \left\{ 
\frac{i}{\hbar} \int_{-\infty}^{t} dt^{\prime}
\textrm{Tr} \left( \hat{\rho}_{\textrm{in}} \hat{H}_{\mathrm{int}}(t^{\prime}) \mathcal{\hat{O}}(t) \right) \right\} \nonumber \\
& & + \frac{1}{\hbar^2} \int_{-\infty}^{t} \int_{-\infty}^{t} dt^{\prime} dt''
\textrm{Tr} \left( \hat{H}_{\mathrm{int}}(t'')
\hat{\rho}_{\textrm{in}} \hat{H}_{\mathrm{int}}(t^{\prime})
\mathcal{\hat{O}}(t) \right).
\end{eqnarray}
Here, $\mathcal{\hat{O}}$ evolves with respect to $\hat{H}_{0} =
\hat{H}_{\textrm{el}}+\hat{H}_{\textrm{rad}}$, i.e., $\mathcal{\hat{O}}(t) =
\textrm{exp}(\tfrac{i}{\hbar}\hat{H}_{0}t)~\mathcal{\hat{O}}
~\textrm{exp}(-\tfrac{i}{\hbar}\hat{H}_{0}t)$, where $\hat{H}_{\textrm{el}}$ and
$\hat{H}_{\textrm{rad}}$ are the Hamiltonians corresponding to the electrons
and the radiation field, respectively. 
If the observable is the intensity at the
detector, the interpretation of the terms on the right-hand side
of Eq.~(\ref{eq3}) is straightforward. The first term describes the
intensity of the incident radiation; the second term describes the
interference between the incident and the scattered radiation; 
and the last term represents the scattered radiation only (see
Fig.~\ref{fig11}).

Now we focus on those terms that are responsible for ultrafast PCI,
i.e., the first and second terms on the right-hand side of
Eq.~(\ref{eq3}). 
Modelling the intensity observable at the detector $\mathcal{\hat{O}}$ 
by the Poynting operator that acts only on the
radiation part~\cite{janmalte2012}, the state of the electronic system after the
scattering event remains undetected. Thus the second term can be
written as
\begin{eqnarray}\label{eq5}
\textrm{Tr} \left( \hat{\rho}_{\textrm{in}}
\hat{H}_{\mathrm{int}}(t^{\prime}) \mathcal{\hat{O}}(t) \right) &
= & \frac{e^{2}}{2~m} \int d^{3}x ~ \textrm{Tr}_{X} \big(
\hat{\rho}^{X}_{\textrm{in}} {\hat{\mathbf{A}}}^2(\mathbf{x},
t^{\prime})
\mathcal{\hat{O}}(t) \big ) \nonumber \\
& & \times \textrm{Tr}_{\textrm{el}} \big(
\hat{\rho}^{\textrm{el}}_{\textrm{in}} \hat{n}(\mathbf{x},
t^{\prime}) \big),
\end{eqnarray}
where $\textrm{Tr}_{X}$ and $\textrm{Tr}_{\textrm{el}}$ are the
traces over radiation and electrons, respectively; and 
$\hat{n}(\mathbf{x}) = \hat{\psi}^{\dagger}(\mathbf{x}) ~ \hat{\psi}(\mathbf{x})$ is the
electron density operator. In this case,
\begin{equation}\label{eq6}
\textrm{Tr}_{\textrm{el}}
\big(\hat{\rho}^{\textrm{el}}_{\textrm{in}} \hat{n}(\mathbf{x},
t^{\prime}) \big) =  \langle \Psi_{\textrm{in}} |
\hat{n}(\mathbf{x}, t^{\prime}) | \Psi_{\textrm{in}} \rangle = \rho(\mathbf{x}, t^{\prime}).
\end{equation}
Here, $\rho(\mathbf{x}, t^{\prime})$ is the IED
of the wavepacket at time $t^{\prime}$. 
At this point, it is interesting to note that the probe pulse inevitably 
changes the wavepacket. But due to 
the interference between the incident and scattered radiation, 
described by Eq.~(\ref{eq5}), eigenstates outside the wavepacket cannot contribute to the signal 
at the detector in the case of near-field imaging [see Eq.~(\ref{eq6})]. This present result is in contrast to the 
results obtained in the case of far-field imaging, where one only observes scattered radiation. 
An analysis of the third term on the right-hand side of Eq.~(\ref{eq3})
demonstrates that the patterns obtained via far-field imaging 
contain the fingerprint of electronic transitions within
the bandwidth of the probe pulse~\cite{dixit2012}.

Now, the x-ray pulse is considered to be quasi-stationary, quasi-monochromatic, and
spatially coherent. Furthermore, we assume that the x-ray pulse duration
is sufficiently short to freeze the dynamics of the
wavepacket, and the pulse propagates along the $y$ axis.
Therefore, using quantum theory as shown in the case of x-ray PCI from a stationary sample~\cite{janmalte2012}, 
we find that the  time-resolved phase contrast image  
is related to the Laplacian of the projected instantaneous
electron density, $\rho_{\perp}(\mathbf{r}_{\perp}, \tau) = \int
dr_{y} \rho(\mathbf{r}, \tau)$ with $\mathbf{r}= (r_{x}, r_{y},
r_{z}), \mathbf{r}_{\perp} = (r_{x}, r_{z})$ and $\tau$ is the pump-probe
delay time. Therefore, the signal at the detector in the case of ultrafast PCI is
\begin{equation}
\label{eq:image}
 I(\mathbf{r}) = I_{\textrm{in}}~\left(  1 - \frac{2\pi r_{e}D}{|\mathbf{k}_{\textrm{in}}|^{2}} \nabla^{2}
\rho_{\perp}(\mathbf{r}_{\perp}, \tau)\right),
\end{equation}
where $I_{\textrm{in}}$ is the signal due to the incident
radiation alone, $\mathbf{k}_{\textrm{in}}$ is the wavevector of the 
incident radiation, $r_{e}$ is the classical
electron radius, and $D$ is the distance of the detector from the
object. 
The key quantity in ultrafast PCI is the Laplacian of the projected IED of the wavepacket.
The experimental parameters for performing ultrafast PCI are subject to certain conditions.
The near-field regime is in general characterized by a Fresnel number $F \gtrsim 1$, which relates
the spatial extension $a$ of the wavepacket with $D$ and the wavelength $\lambda$
of the incident radiation via $F=\tfrac{a^{2}}{\lambda D}$. Quasi-monochromaticity of the pulse is ensured by 
a sufficiently long coherence time;
however, $\tau_{\mathrm{tc}}$ has to be short with respect to the propagation time of the radiation to the detector,
$ \frac{\lambda}{2\pi \,c} \ll \tau_{\mathrm{tc}} \ll \frac{D}{c}$, $c$ denoting the speed of light.

The two-dimensional (2D) projection of the IED, $\rho_{\perp}$, can be retrieved by applying a suitable
Poisson solver algorithm to a single recorded image.
The full three-dimensional (3D) tomographic
distribution of the IED can be reconstructed by recording several
2D projected images for many different projection angles.
Several tomographic methods to reconstruct the 3D distribution
have been developed, as reviewed by Burvall {\em et al.}~\cite{burvall2011}.
A direct method for reconstructing the 3D IED, without intermediate
reconstruction of the 2D projected IED $\rho_\perp$, was presented in Ref.~\cite{bronnikov2002}.
By taking images of the object rotated around the $z$ axis about an angle
$\theta$, described by the density $\rho_\theta$,
one obtains the data sets $g_\theta=\nabla^2{\rho_{\theta}}_\perp$ from Eq.~\eqref{eq:image}.
The reconstruction algorithm has the form of a filtered backprojection.
By the Fourier projection theorem \cite{kak2001principles} the Fourier transform of a data function $g_\theta$
corresponds to a rotated plane in Fourier space of the 3D object.
The 3D object can be reconstructed by applying the filter function $Q(k_{x_\theta},k_z)=|k_{x_\theta}|/(k_{x_\theta}^2+k_z^2)$
in Fourier space and integrating over all rotation angles,
\begin{equation}
 \rho(\mathbf{r}, \tau) = \frac{1}{4\pi^2} \int_0^\pi d\theta \mathcal{F}^{-1}[ Q\mathcal{F}[g_\theta] ](x_\theta, z)\,,
\end{equation}
where $x_\theta = x\cos\theta+y\sin\theta$ and $\mathcal{F}$ denotes the 2D Fourier transform.
The 3D Laplacian of the IED, $\nabla^2\rho(\mathbf{r}, \tau)$, can be obtained by applying the filter
$Q(k_{x_\theta}, k_z)=|k_{x_\theta}|$ in the filtered backprojection.
Therefore, ultrafast PCI provides the full 3D Laplacian of the IED, which
highlights internal and external boundaries of the wavepacket e.g., zero flux boundaries---a very useful and informative quantity
according to Bader's theory of atoms in molecules, where
the Laplacian of the electronic density is used to provide detailed
information about the complex bonding and topology of the charge
distributions in molecules~\cite{bader1981, bader1991}.

In order to illustrate the generality of our proposed method for imaging the IED of the wavepacket, 
we apply ultrafast PCI to the same electronic wavepacket as
was considered by Dixit {\em et al.}~\cite{dixit2012}.
In this case, the electronic wavepacket is prepared as
a coherent superposition of the $3d$ and $4f$ eigenstates of atomic hydrogen, 
each eigenstate having a population of 50\% and the projection of orbital angular momentum being 
equal to zero. The oscillation period of the wavepacket is T = 6.25 fs, which is inversely related to the
energy spacing between the $3d$ and $4f$ eigenstates (0.66 eV). It is important to emphasize that 
the considered wavepacket lies in the energy and timescale of the valence electronic motion  
in more complex molecular and biological systems~\cite{breidbach2003, kuleff2005, remacle, scholes}. 
Moreover, in many-electron systems, where only few electrons get excited and participate in the
formation of an electronic wavepacket,
the electronic density of the system can be decomposed into a time-dependent
density of excited electrons and a static density of the stationary electrons.

\begin{figure*}
\includegraphics[width=12cm]{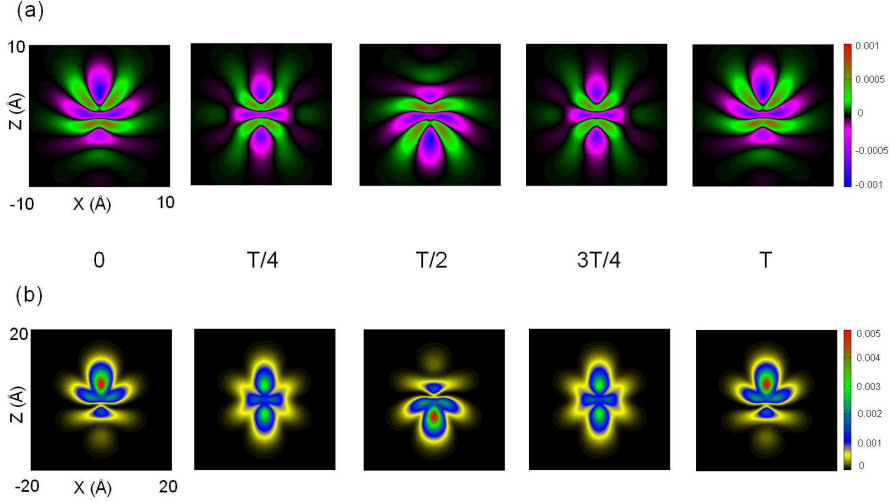}
\caption{(color online). Phase contrast images and projected
instantaneous electron density of the wavepacket. (a) Laplacian of
the projected instantaneous electron density, and (b) the
instantaneous electron density integrated along the direction of
propagation of the incident \mbox{x rays} ($y$ axis) at
pump-probe delay times 0, T/4, T/2, 3T/4, and T, where the
oscillation period of the electronic wavepacket is T = 6.25 fs.}
\label{fig1}
\end{figure*}

Figure~\ref{fig1}(a) shows phase contrast images calculated with the
Laplacian of the projected IED,
$\nabla^{2} \rho_{\perp}$ as a function
of the delay time at times 0, T/4, T/2, 3T/4, and T.
Figure~\ref{fig1}(b) shows the projected IED, $\rho_{\perp}$, of the wavepacket,
where the electronic charge distribution periodically undergoes
localization to delocalization from positive to negative $z$ axis
within 6.25 fs. The spatial
extension of $\rho_{\perp}$ is 14--17 \AA~along 
the $z$ axis and 7.5--9 \AA~along the $x$ axis. 
As the electronic charge distribution oscillates from positive to negative 
$z$ axis and returns back to its original position, the patterns obtained by 
ultrafast 
PCI follow the same trend of the oscillation and provide the correct time period of the oscillation.  
At delay times T/4 and 3T/4, the charge distributions are completely delocalized
and identical to each other, and the phase contrast images are also 
identical (see Fig.~\ref{fig1}). The present results are fundamentally different 
from the scattering patterns obtained via far-field scattering, where
the patterns are completely different at delay times T/4 and 3T/4, whereas the electronic
charge distributions are identical at those times~\cite{dixit2012}.
Similarly, the charge distributions are completely different at times 0 and T/2, 
which is also reflected in the phase contrast images at those delay times, whereas the far-field scattering
patterns are identical at times
0 and T/2~\cite{dixit2012}. Clearly, the images obtained with ultrafast 
PCI are directly related to the IED of the wavepacket.

The patterns obtained via ultrafast PCI are rich in
information. Figure~\ref{fig1}(a) shows more structures than the
corresponding $\rho_{\perp}$ shown in
Fig.~\ref{fig1}(b). It is important to notice that the images
have both positive and negative lobes, whereas the densities have
only positive lobes. 
The $3d$ and $4f$ eigenstates in the wavepacket have no radial
nodes, they only have angular nodes. The PCI images show
the curvature of these angular nodes in the
wavepacket. The electronic charge distribution is locally
concentrated where $\nabla^{2} \rho_{\perp} < 0$ and depleted where
$\nabla^{2} \rho_{\perp} > 0$. Therefore, these images provide
the topology of the electronic charge distribution with detailed
information about the nodal structure of the wavepacket. 

In conclusion, we have demonstrated that the IED of the wavepacket can be imaged via 
ultrafast PCI. 
Therefore, ultrafast PCI can provide direct information about the dynamical changes in the 
spatial electron probability distribution at different instants of time.
Morever, ultrafast PCI provides the Laplacian of the IED, which reveals the internal 
structures of the wavepacket through local variations in the IED. 
Ultrafast PCI does not rely on any assumptions such as the single-active-electron approximation~\cite{itatani2004, blaga2012}. 
The technique is therefore applicable to large and strongly correlated  
systems and ultrafast PCI would be very powerful
for imaging electronic quantum motion in real-time and
real-space with potentially unprecedented spatio-temporal resolution.
Ultrafast PCI may be expected to shed light on 
non-equilibrium quantum motion, for example, 
in peptides and biological systems~\cite{breidbach2003, kuleff2005, remacle}.
Therefore, with the tremendous technological advancement in x-ray imaging
technologies~\cite{sakdinawat2010, gruner2012}, one might expect
ultrafast PCI measurements to be feasible in the future.

\begin{acknowledgments}
We thank Sang-Kil Son for careful reading of the manuscript.
\end{acknowledgments}

\end{document}